\documentstyle[12pt,epsf]{article}
\newcommand{\be}{\begin{equation}}
\newcommand{\ee}{\end{equation}}
\newcommand{\bea}{\begin{eqnarray}}
\newcommand{\eea}{\end{eqnarray}}
\newcommand{\nn}{\nonumber}
\newcommand{\Seff}{S_{\mbox{eff}}}

\newcommand{\meff}{m_{\mbox{eff}}}
\newcommand{\dsla}{\partial \hspace{-2mm}/}
\newcommand{\psla}{p \hspace{-2mm}/}
\newcommand{\ksla}{k \hspace{-2mm}/}
\newcommand{\Asla}{A \hspace{-2mm}/}
\begin{document}
\vspace{-1cm}
\noindent
\begin{flushright}
KANAZAWA-99-12\\
KUCP-0140
\end{flushright}
\vspace{20mm}
\begin{center}
{\Large \bf
Analysis of the Wilsonian Effective Potentials
\vspace{2mm}\\
in Dynamical Chiral Symmetry Breaking
}
\vspace*{15mm}\\
Ken-Ichi Aoki
%\footnote{e-mail address: aoki@hep.s.kanazawa-u.ac.jp}
, Keiichi Morikawa
%\footnote{e-mail address: morikawa@hep.s.kanazawa-u.ac.jp}
, Jun-Ichi Sumi$^\dagger$
%\footnote{e-mail address: sumi@hep.s.kanazawa-u.ac.jp}
,Haruhiko Terao
%\footnote{e-mail address: terao@hep.s.kanazawa-u.ac.jp}
\\
and Masashi Tomoyose
%\footnote{e-mail address: tomoyose@hep.s.kanazawa-u.ac.jp}
\vspace*{3mm}\\
Institute for Theoretical Physics, Kanazawa University\\
Kanazawa 920--1192, Japan
\vspace*{3mm}\\
$^\dagger$Department of Fundamental Sciences,\\
Faculty of Ingegrated Human Studies, Kyoto University,\\
Kyoto 606-8501, Japan
\vspace*{40mm}\\
{\large \bf Abstract}
\end{center}

%%%%%%%%% ABSTRUCT %%%%%%%%%%
The non-perturbative renormalization group equation for the Wilsonian 
effective potential is given in a certain simple approximation scheme 
in order to study chiral symmetry breaking phenomena dynamically
induced by strong gauge interactions.
The evolving effective potential is found to be non-analytic in
infrared, which indicates spontaneous generation of the fermion mass.
It is also shown that the renormalization group equation gives the 
identical effective fermion mass with that obtained by solving
the Schwinger-Dyson equation in the (improved) ladder approximation.
Moreover introduction of the collective field corresponding to 
the fermion composite into the theory space is found to offer an 
efficient method to evaluate the order parameters; the dynamical mass
and the chiral condensate.
The relation between the renormalization group equation incorporating
the collective field and the Schwinger-Dyson equation is also 
clarified. 

%%%%%%%%%% INTRODUCTION %%%%%%%%%%%
\setcounter{footnote}{0}
\newpage
\pagestyle{plain}
\pagenumbering{arabic}
\noindent
{\large \bf 1. Introduction}
\vspace*{2mm}

It has been a very important subject to understand the
non-perturbative dynamics of the quantum field theories, especially
the strongly coupled gauge theories like QCD.
One of the most outstanding features of the non-perturbative dynamics
of QCD is spontaneous breakdown of the chiral symmetry. 
So far, the approach by means of the Schwinger-Dyson equations (SDE)
has been extensively studied \cite{SDmain,SDfourfermi,SDphase} 
and been applied not only to the strongly coupled QED and QCD 
\cite{SDimproved,SDQCD}, but also to 
the models beyond the satandard model \cite{SDmodel,topcondensation}.
In this approach the so-called (improved) ladder approximation 
\cite{SDimproved} has been commonly applied to solve the SDEs.
It should be noted that the SDEs offers us surprisingly good results 
on the chiral symmetry breaking phenomena in QCD in spite of the fact
that the approximation scheme is fairly crude \cite{SDQCD}.
However it has been also known that this SD approach suffers from some
problems: difficulties in improvement of the approximation
\cite{SDbeyond}, 
large gauge dependence in the ladder approximation \cite{SDdifficulty}, 
and so on.
Therefore it would be desirable to develop other non-perturbative 
methods so as to overcome such difficulties, for example, in order
to perform gauge independent study of the chiral order parameters
in QCD dynamics.

In the previous papers \cite{ccbqed,RGfourfermi}, we proposed the Wilson 
renormalization group equation (RGE) \cite{WK} for the system of a 
massless fermion coupled by gauge interaction, with which we can
study the critical behavior of the chiral symmetry in a remarkably 
simple manner. 
In this RG approach it is enough to calculate the beta functions 
for the multi-fermi couplings in order to evaluate the critical 
couplings, phase boundary and the anomalous dimensions of the 
fermion composites near the critical points. Specially the RG flows
of the four-fermi coupling play an essential role to show this critical
behavior. 
The procedure is quite straightforward in contrast to the calculations
using the SDEs.
It should be noted that the set of beta functions obtained in a 
certain approximation framework turns out to be free from gauge 
dependence \cite{ccbqed}. 
Moreover, if we approximate these RGEs further by restricting the 
radiative corrections to a certain limited type, which we called the 
``ladder-like'' corrections in \cite{RGfourfermi}, then it gives the same 
critical dynamics as obtained by solving the ladder SDE. 
Therefore the above mentioned gauge independent analysis indeed
offer us an improvement of the ladder SDEs.
Thus, it may be said that this Wilson RG approach is more powerful
than the SD approach as long as the critical dynamics is concerned.

Howeve the beta functions for the multi-fermi couplings do not tell
us the order parameters in spontaneous chiral symmetry breaking.
Therefore it is not known immediately in which phase the chiral symmetry 
is broken, even after the critical behavior is clarified by the RGEs.
Actually the chiral symmetry breaking itself might look somehow
puzzling in the framework of the Wilson RG. Because the Wilsonian 
effective action keeps chiral invariant manifestly, and therefore 
the fermion mass seems to have no chance to appear in it.
On the other hand, the order parameters such as the dynamical mass of
quarks, the quark condensation, the $\pi$-on decay constant, etc. 
are the relevant physical quantities to be evaluated in QCD-like gauge 
theories, which are of our main concern.
Thus it is an interesting and important problem to study not only the
critical behavior but also the chiral order parameters in this RG 
framework. 

In this paper we discuss how the order parameters may be evaluated in the
Wilson RG framework. 
For the sake of simplicity we adopt the simple approximation to take 
only the ``ladder like'' corrections into account and propose a 
non-perturbative RGE in terms of the Wilsonian effective potential.
Then the dynamical fermion mass is found to appear through 
non-analitycity of the Wilsonian effective potential in infrared 
instead of an explicit mass term. 
Also it will be shown that the RGE gives the same effective mass 
as the ladder SDE does in the entire region of the coupling space. 
Through this analysis the structure of the ladder SDE is also
clarified in the point of view of the non-perturbative RG.
Thus any improvement of this approximation in the RG
framework exceeds the analysis done by the ladder SDE so far. 
However the purpose of this paper is to discuss the method to evaluate
the order parameters in the Wilson RG framework and to make clear the
relation to the SDE, not to improve results for the order parameters 
obtained so far.
Study of the order parameters beyond the level of the SD approach, 
specially the gauge independent analysis, is currently under way.

This paper is organized as follows.
Section 2 is devoted to a brief review of the SDE in the so-called 
improved ladder approximation, which will be disscussed in relation 
to the RGE in the later sections. 
In section 3 the non-perturbative RGE for the Wilsonian effective 
potential is proposed. 
The spontaneous chiral symmetry breaking will be examined by solving 
the partial differential equation for the effective potential. 
It will be seen that in the chiral symmetry broken phase the 
non-analyticity comes out in the infrared effective potential. 
This singular behavior is found to indicate the dynamical generation
of the fermion mass.
The relation between the RGE and the SDE is clarified by an analytical
method in section 4. 
Through this observation the improved ladder approximation used in the 
SD approach will be understood from the point of view of Wilson RG. 
However the non-analytic behavior of the Wilsonian effective potential
makes it rather difficult to evaluate the order parameters.
In section 5 we are going to consider the practical method to 
calculate the order parameters in this RG framework.
In order to circumvent this problem we will propose a method 
incorporating the ``collective field'' corresponding to the 
composite operator of fermions in the theory space. 
This method will be found to enable us to perform the calculation very 
effectively compared with the former RGEs.
Lastly the results of the effective mass of quarks and the chiral 
condensate in massless QCD will be presented using the RGEs in 
comparison with the results obtained in the ladder SD approach. 
In section 6 we summarize our conclusions and also discuss the further 
issues along the lines considered in this paper.

\vspace*{4mm}
\noindent
{\large \bf 2. The ladder Schwinger-Dyson equation}
\vspace*{2mm}

First let us consider the massless QED extended so as to include
the chiral invariant 4-fermi intreactions; namely the gauged
Namubu-Jona-Lasinio model \cite{SDfourfermi,SDphase}, 
which is given by the bare action
\be
S_0 = \int d^4x~\bar{\psi}(i\dsla - g \Asla)\psi
-\frac{1}{4}F_{\mu\nu}^2
+\frac{G}{2\Lambda_0^2}\left[
(\bar{\psi}\psi)^2+(\bar{\psi}i\gamma_5\psi)^2
\right].
\ee
This should be regarded as the action of the effective theory at scale
$\Lambda_0$, which is the overall cut-off of this model. 
The 2-point function of the fermion $S(p)$, which is of our present 
interest in order to see chiral symmetry breaking, satisfies the SDE 
written in terms of the full photon propagator $D_{\mu\nu}(p)$, the 
full vertex function $g\Gamma_{\mu}(p, q)$ and $S(p)$. 
However the SDE for the vertex cannot be given a closed form using
only $S$, $D_{\mu\nu}$ and $g\Gamma_{\mu}$. 
Therefore it is inevitable to approximate the SDE in any practical 
analyses.

In the ladder approximation, which has been mostly used, the photon 
propagator $D_{\mu\nu}$ as well as the vertex function $g\Gamma_{\mu}$ 
are simply replaced by the tree level ones;
\bea
& &g\Gamma_{\mu}(p, q)=g\gamma_{\mu} \nonumber \\
& &iD_{\mu\nu}(p)=\frac{1}{p^2}\left(
g_{\mu\nu}-\frac{p_{\mu}p_{\nu}}{p^2}
\right),
\eea
where the propagator is given in the Landau gauge. 
Actually the results obtained in this approximation have been known to 
depend on the gauge choice considerably \cite{SDdifficulty}. 
However the SDE in the ladder approximation is supposed to be the best 
in the Landau gauge, since the Ward identity $Z_1=Z_2$ is maintained 
only in this gauge. Indeed $Z_1=1$ by neglecting corrections to the 
vertex and also $Z_2=1$ as is seen from the form of the 2-point 
function in the Landau gauge,
\be
iS^{-1}(p)=\psla - \Sigma(p),
\ee
where $\Sigma(p)$ is the fermion self-energy i.e. the mass function.

The SDE for the mass function $\Sigma$ is found to be given by the
integral equation (in Euclidean)
\be
\Sigma(k)=\int_{|p|<\Lambda_0} \frac{d^4 p}{(2\pi)^4}
\left[ \frac{G}{\Lambda_0^2}+\frac{3g^2}{4(k-p)^2} \right]
\mbox{tr}\left( \frac{1}{i\psla + \Sigma(p)} \right),
\ee
which was first studied by Bardeen et. al. 
\cite{SDfourfermi}. It is seen from this
equation that the mass function may be evaluated as sum of bubble diagrams
of the fermion loops which contain the ``ladder'' corrections of photon
exchange. The spontaneous chiral symmetry breaking is indicated by 
appearance of a non-trivial solution of the eq.(4). 
The structure of the critical line in the space of the bare couplings 
$(G, \alpha \equiv g^2/4\pi^2)$ has been clarified \cite{SDphase} 
and the symmetry broken phase was found to be given by
\be
\left\{
\begin{array}{rlcc}
G > & \pi^2\left( 1 + \sqrt{1-\alpha/\alpha_{\mbox{cr}}}\right)^2
& \mbox{for} & \alpha<\alpha_{\mbox{cr}}, \\
\alpha > & \alpha_{\mbox{cr}}
& \mbox{for} & G < \pi^2,
\end{array}
\right.
\ee
where $\alpha_{\mbox{cr}}$ denotes the critical gauge coupling and is
found to be $\alpha_{\mbox{cr}}=\pi/3$ \cite{SDmain}. 

The ladder approximation mentioned above, however, is so crude that
the radiative corrections to the gauge coupling are totally ignored.
Therefore we cannot apply the SDE (4) directly to the gauge theories
such as QCD in order to obtain any physical results.
Miransky and Higashijima \cite{SDimproved} have invented independently
the scheme to incorporate the correction to the gauge coupling to the
SDE effectively, which is called the improved ladder approximation. 
This approximation is to simply replace the vertex $g\gamma_{\mu}$ by
\be
g\Gamma_{\mu}=\bar{g}\left(\mbox{max}(p^2, k^2)\right)
\gamma_{\mu}T^a,
\ee
where $p$ and $k$ are the Euclidean momentum carried by the fermions
attached to the vertex, $\bar{g}(p^2)$ is the running gauge coupling
constant renormalized at the scale of $p$. 
Also $T^a$ denotes the color group representation of the fermion. 
The choice of the momentum dependence in (6) is just for the practical 
reason to make analytical study easier.
Thus the SDE in the improved ladder approximation is written after
the angle integration as
\be
\Sigma(k)=\frac{1}{2\pi^2}\int_0^{\Lambda_0} dp ~p^3 
\left[G+\left(
\frac{\lambda(k)}{k^2}\theta(k-p)+
\frac{\lambda(p)}{p^2}\theta(p-k)
\right)\right]
\frac{\Sigma(p)}{p^2 + \Sigma^2(p)}.
\ee
Here we introduced the running gauge coupling defined by
\be
\lambda(p)=\frac{3}{4}C_2(R)\bar{g}^2(p),
\ee
where $C_2(R)$ is the second Casimir of the color representation $R$. 
We may use the running gauge coupling evaluated by perturbation
in the SDE, provided an infrared cutoff of running is performed to
$\lambda(p)$ in order to avoid the divergent pole. Fortunately it has 
been found that the physical quantities are almost independent of this 
infrared cutoff \cite{SDQCD}.

Indeed the scheme in the improved ladder approximation makes it possible
to study QCD-like gauge theories. 
Besides it has been known that the SDE (and also the Bethe-Salpeter 
equations) in this scheme leads to the order parameters for dynamical 
chiral symmetry breaking in QCD in a rather good accuracy \cite{SDQCD}. 
However this method to use the running gauge coupling deviates from
the scheme of SDE, and therefore, it cannot be regarded as a
systematic improvement of the approximation. 
Hence we cannot help but stopping more or less in this level of 
analyses. 
However it will be seen later that this manipulation is understood 
as a quite natural improvement of the approximation in the framework 
of non-perturbative renormalization group rather than in this SD 
approach.

\vspace*{4mm}
\noindent
{\large \bf 3. The Wilsonian effective potential and the RG equations}
\vspace*{2mm}

There have been known several formulations of the non-perturbative
RG or the so-called exact renormalization group equations 
\cite{WK,WH,exactRG}. 
In this paper we discuss the dynamical chiral symmetry breaking 
phenomena only in the framework of the Wegner-Houghton RGE 
\cite{WH}
for our present purpose.
This RGE gives the variation of the so-called Wilsonian effective 
action defined by sharp cutoff under the infinitesimal shift 
of the cutoff scale. 
As is seen later this sharp cutoff scheme makes it easier to clarify 
the relation between the non-perturbative RGE's and the SDE's, though 
the physical consequencies should not depend on the choice of cutoff 
scheme of course.

Before discussing the RGE for the system considered in the previous
section, let us mention briefly the general formulation of the
Wegner-Houghton RGE.  
If we devide the freedom of the quantum field $\phi(p)$ into the
higher momentum modes with $|p|>\Lambda$ and the lower momentum modes 
with $|p|<\Lambda$ by introducing the cutoff scale $\Lambda$ in the 
Euclidean formalism, then the Wilsonian effective action at this 
scale, $\Seff[\phi;\Lambda]$, may be defined by integrating out the 
higher frequency modes in the partition function. 
Namely
\be
Z=
\int \prod_{|p|<\Lambda_0}d\phi(p)~e^{-S_0[\phi;\Lambda_0]}
=\int \prod_{|p|<\Lambda}d\phi(p)~e^{-\Seff[\phi;\Lambda]},
\ee
where $S_0$ denotes the bare action with the bare cutoff $\Lambda_0$.
This effective action contains the general operators consistent
with the original symmetries of the bare action, for example the
chiral symmetry of our concern. 
Interestingly enough the infinitesimal variation of the effective 
action with respect to the cutoff $\Lambda$ may be evaluated exactly. 
After taking account of the scale transformation of dimensionful 
variables the Wegner-Houghton RGE is found to be
\bea
& &\frac{\partial \Seff}{\partial t}
=d\Seff-
\int \frac{d^dp}{(2\pi)^d}~\phi^i_p\left(
\frac{2-d-\eta^i}{2}-p^{\mu}\frac{\partial'}{\partial p^{\mu}}
\right)
\frac{\delta\Seff}{\delta \phi^i_p} \nonumber \\
& &\hspace*{12mm}
-\frac{1}{2}\int \frac{d^dp}{(2\pi)^d}\delta(|p|-1) 
\left\{
\frac{\delta \Seff}{\delta\phi^i_p}
\left(
\frac{\delta^2 \Seff}{\delta\phi^i_p\delta\phi^j_{-p}}
\right)^{-1}
\frac{\delta \Seff}{\delta\phi^j_{-p}}
-\mbox{\rm str}\ln 
\left(
\frac{\delta^2 \Seff}{\delta\phi^i_p\delta\phi^j_{-p}}
\right)
\right\},
\eea
where $t=\ln(\Lambda_0/\Lambda)$ is introduced as the scale parameter.
The 1st line of the RGE represents nothing but the scaling of the effective
action. $\eta^i/2$ denotes the anomalous dimension of the field $\phi^i$.
While the 2nd line comes from the radiative corrections which
correspond to the tree and the one loop Feynman diagrams including only 
the propagators with the momentum of the scale $\Lambda$. However this
RGE is a rather formal object, though it has been derived exactly.
It is inevitable to simplify this RGE by applying some approximations
in practical analyses.

Now we shall consider the non-perturbative RGE for the system given by the 
bare action (1) in the approximation scheme, which will be found to 
correspond to the ladder approximation applied to the SDE.
In the process to derive the RGE, the scheme of this approximation
will be defined. 
First we adopt the so-called local potential approximation (LPA)
\cite{WH,LPA}, in which
the radiative corrections to any operators containing
derivatives are ignored. Therefore solely the potential part
of $\Seff$ may be evolved with respect to the cutoff scale. 
It should be noted that the wave function renormalizations, therefore 
the anomalous dimensions also, are ignored in this scheme. 
First let us also ignore the corrections to the operators including 
the gauge field. 
In the Wilson RG framework the gauge invariance is necessarily lost by 
introducing cutoff, though the chiral invariance is maintained. 
This problem makes it rather complicated to deal with gauge theories 
generally. 
Recently the formulations of the Wilson RG utilizing the
Slavnov-Taylor identities have been developed and been examined in the 
application to the Yang-Milles theories \cite{MSTI}. 
However in the present approximation we avoid this problem just by 
ignoring such corrections. 
One the other hand we do not get the radiative corrections to the
gauge interaction at the price of this approximation, which is in the 
same level as of the ladder SDE. 
Treatment of the running gauge coupling will be discussed later.

As the result of the approximation disscussed so far, the form of 
the Wilsonian effective action at scale $\Lambda$ is restricted to 
\be
\Seff[\psi, \bar{\psi}, A_{\mu}; \Lambda]=
\int d^4x~ \bar{\psi}(\dsla + g \Asla)\psi + V(\psi, \bar{\psi}; \Lambda)
+\frac{1}{4}F_{\mu\nu}^2 + \frac{1}{2\alpha}(\partial_{\mu}A_{\mu})^2,
\ee
where $V$ is the general potential invariant under the chiral symmetry.
The last term in (11) is the gauge fixing term and we also adopt the 
Landau gauge hereafter. 
The potential $V$ may be written down as a polynomial composed 
of the following parity and chiral invariant operators, which are 
mutually independent;
\bea
& &~~{\cal O}_1 =
(\bar{\psi}\psi)^2+(\bar{\psi}i\gamma_{5}\psi)^2
= -\frac{1}{2}\left\{
(\bar{\psi}\gamma_{\mu}\psi)^2-(\bar{\psi}\gamma_{5}\gamma_{\mu}\psi)^2
\right\}, \nonumber \\
& &~~{\cal O}_2 =
(\bar{\psi}\gamma_{\mu}\psi)^2+(\bar{\psi}\gamma_{5}\gamma_{\mu}\psi)^2,\\
& &~~{\cal O}_3 =
\left\{
(\bar{\psi}\gamma_{\mu}\psi)(\bar{\psi}\gamma_{5}\gamma_{\mu}\psi)
\right\}^2. \nonumber
\eea

Now the infinitesimal variation of the effective potential 
$V(\psi, \bar{\psi}; \Lambda)$ under the shift of the cutoff 
$\Lambda \rightarrow \Lambda-\delta\Lambda$ 
is given by
\bea
\delta V(\psi, \bar{\psi}; \Lambda)
&=&
\int_{\Lambda-\delta\Lambda < |p| < \Lambda}\frac{d^4 p}{(2\pi)^4}
\mbox{str}\ln M \nn \\
&=&
\int_{\Lambda-\delta\Lambda < |p| < \Lambda}\frac{d^4 p}{(2\pi)^4}
\left[
\mbox{tr}\ln M_{BB}-\mbox{tr}\ln (M_{FF}-M_{FB}M_{BB}^{-1}M_{BF})
\right],
\eea
where $M$ is the matrix given by 
\be
M=
\left(
\begin{array}{cc}
M_{BB} & M_{BF} \\
M_{FB} & M_{FF} 
\end{array}
\right)
=
\left(
\begin{array}{ccc}
D_{\mu\nu}^{-1} & g\bar{\psi}\gamma_{\mu} & -g\gamma_{\mu}\psi \\
-g\bar{\psi}\gamma_{\nu} & 
\frac{\delta^2 V}{\delta \psi\delta \psi} &
-i\psla^{T}+\frac{\delta^2 V}{\delta \psi\delta \bar{\psi}} \\
g\gamma_{\nu}\psi & 
-i\psla+\frac{\delta^2 V}{\delta \bar{\psi}\delta \psi} &
\frac{\delta^2 V}{\delta \bar{\psi}\delta \bar{\psi}} 
\end{array}
\right).
\ee
Here $D_{\mu\nu}=M_{BB}^{-1}$ denotes the photon propagator in the
Landau gauge. Note that the combination of $M_{FB}M_{BB}^{-1}M_{BF}$ 
in eq.(13) represents the effective 4-fermi interactions induced by 
one photon exchange.

In the previous paper \cite{RGfourfermi} we examined the critical 
dynamics of 
chiral symmetry breaking in this level of approximation. 
However it was found that this approximation is still better than the 
ladder approximation in the SD approach as far as the critical 
dynamics is concerned. 
In other words this correction contains the contributions from 
the ``non-ladder'' diagrams not only from the ``ladder'' diagrams. 
Here we shall simply extract the ``ladder-like'' corrections only by 
imposing further restriction to the effective potential. 
We consider the effective potential $V$ composed of only the operator 
${\cal O}_1$ given in (12), and pick up only the parts propotional to 
$1$ or $\gamma_{5}$ in the spinor structure from the vertices
appearing in $N_{FF}\equiv M_{FF}-M_{FB}M_{BB}^{-1}M_{BF}$. 
After performing the fierz transformation the resultant matrix is 
simply reduced to be 
\be
N_{FF}=
\left(
\begin{array}{cc}
0 & -i\psla^{T}-(\sigma+i\gamma_{5}\pi)(V_{\rho}-\frac{3g^2}{4p^2})\\
-i\psla + (\sigma+i\gamma_{5}\pi)(V_{\rho}-\frac{3g^2}{4p^2}) & 0 
\end{array}
\right),
\ee
where the notations 
$\sigma=\bar{\psi}\psi$, $\pi=\bar{\psi}i\gamma_{5}\psi$
and $\rho=(\sigma^2+\pi^2)/2$ are introduced. 
\footnote{
This matrix (15) in LPA is evaluated simply by setting the lower
momentum modes to zero modes, however there arises a subtle problem 
related with their Grassmann nature. 
The effective potential evaluated by the exact zero modes must 
terminate at a certain finite order, unless we consider the large
$N_f$ limit. However such zero mode parts do not represent
the bulk structure of the general Green functions of fermions. 
Here we suppose that the correction to the effective potential
is evaluated in small momentum limit with keeping $\delta \Lambda$
finite rather than by the exact zero modes, and that the effective 
potential $V$ is given by an infinite order of polynomial in terms of 
$\sigma=\bar{\psi}(-p)\psi(p)$ and
$\pi=\bar{\psi}(-p)i\gamma_{5}\psi(p)$. 
The sharp cutoff limit is taken by $\delta \Lambda > |p| \rightarrow 0$.
}
Thus we obtain the non-perturbative RGE in the ``ladder''
approximation as
\be
\frac{\partial V}{\partial t}=
4V-6\rho V_{\rho} - \frac{1}{4\pi^2}\ln
\left[
1 + 2\rho(V_{\rho}-3\pi\alpha)^2
\right],
\ee
after taking account of the canonical scaling.
Indeed it will be shown that analysis of this RGE results in
the identical effective mass, which is defined shortly, to the
solution of the ladder SDE (4).  

Before discussing the spontaneous chiral symmetry breaking, let us 
briefly mention the critical surface derived from this RGE. If we
expand the effective potential $V(\rho)$ into a polynomial,
\be
V(\rho; t)=
- G(t)\rho + \frac{1}{2}G_8(t)\rho^2 
+ \frac{1}{3!}G_{12}(t)\rho^3 + \cdots,
\ee
and substitute this into the RGE, then the beta functions for the 
couplings are easily obtained. For the 4-fermi coupling $G$ we find
\be
\frac{d G}{dt}=-2G+\frac{1}{2\pi^2}\left(G+\frac{3}{\pi}\alpha\right)^2.
\ee
Here we note that the higher order couplings are not involved in the
evolution of the 4-fermi coupling due to 1-loop nature of the 
non-perturbative RGE. This RGE (18) shows the fixed points (line) at
\be
G^* (\alpha) = \pi^2\left(1\pm\sqrt{1-\alpha/\alpha_{\mbox{cr}}}\right)^2,
\ee
where $+$ ($-$) sign is for the UV (IR) fixed points respectively. 
Thus the phase boundary, which is shown in Fig.1, is found to just 
coincide to the result of the ladder SDE given in eq.(5). 
The anomalous dimension of the four-fermi operator is immediately 
obtained from this RGE \cite{RGfourfermi}, 
which is also shown to agree with that derived from the SDE (4) 
\cite{SDphase}.
However it is not obvious whether the chiral symmetry is truly broken
spontaneously in the upper region of the phase boundary in Fig.1, 
since the order parameters of chiral symmetry are not obtained in such 
RG analyses.

\begin{figure}[hbt]
\epsfxsize=0.66\textwidth
\begin{center}
\leavevmode
\epsffile{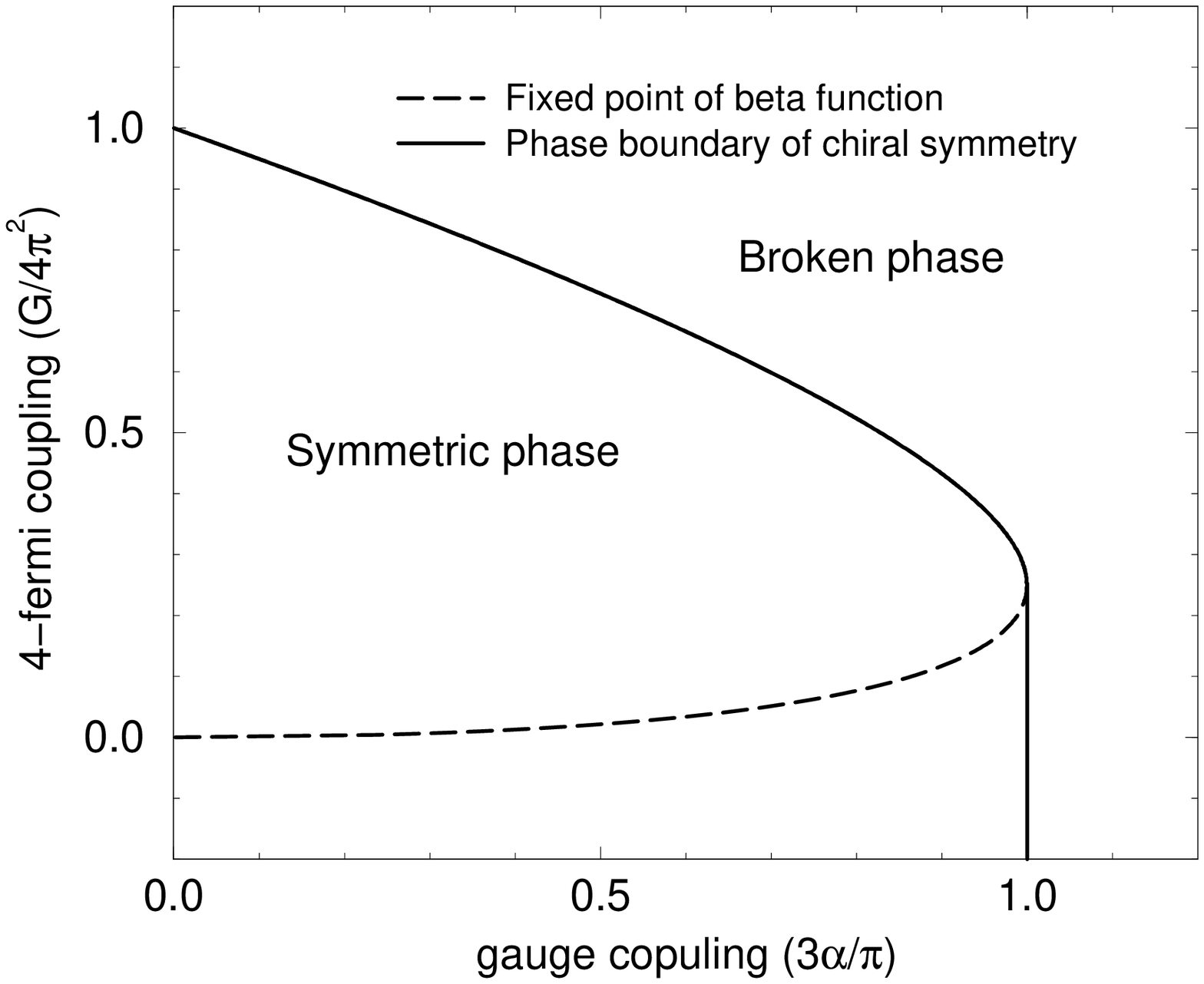}
\\
\parbox{130mm}{
Fig.1: The phase diagram of the chiral symmetry and the fixed line
obtained in the ladder approximation.
}
\end{center}
\end{figure}

Let us consider the effective mass of the fermion, which is the
direct result obtained by solving the SDE. In the RG framework
the Wilsonian effective action keeps it's chiral invariance under
evolution. Therefore the effective mass cannot appear as the mass
term of this action. On the other hand the effective mass $\meff$ 
enters into the loop correction may be read from the effective 
interaction given in eq.(15) as
\be
\meff 
= \lim_{\Lambda \rightarrow 0}
\left. \sigma\frac{\partial V}{\partial \rho}\right|_{\sigma=\pi=0}
= \lim_{\Lambda \rightarrow 0}
\left. \frac{\partial V}{\partial \sigma}\right|_{\sigma=\pi=0}.
\ee
If we are allowed to expand the effective potential into a power series 
as in (17), namely, if the potential is a regular function of
$\rho$ at the origin, 
then this effective mass cannot help but vanishing. On the other hand 
it is found that each coupling actually keeps growing very rapidly 
and eventually diverges in the broken phase. 
\footnote{In practice all the couplings diverge at a certain finite 
scale. This divergence occuring at a finite scale, rather than in the 
infrared limit, is supposed to be caused by the naive approximation 
applied here.}
This is nothing but infrared singularity caused by massless particles. 
Note that each beta function is given through one loop diagrams with the 
massless fermion even in the chirally broken phase. 

The above observation would imply that the expansion by the series 
of operators becomes invalid in the infrared.
Hence we examined to numerically solve the partial differential 
equation given by eq.(16).
In Fig.2 evolution of the effective potential to infrared is presented
in the case of bare couplings $G=0, \alpha=1.5\alpha_{\mbox{cr}}$. 
Actually it is seen that the effective potential turns out to be
non-analytic at the origin ($\sigma=\pi=0$). 
Therefore the effective mass given by eq.(20) is not 
well-defined. 
If we introduce the small bare mass of fermion, then the dynamical 
mass for the massless theory may be defined by the vanishing limit 
and is found to get a finite value.
Anyway the dynamical generation of fermion mass appears through the
non-analytic behavior of the Wilsonian effective potential in infrared.
Actual evaluation of the effective mass and also the chiral condensate
will be discussed in section 5.

\begin{figure}[hbt]
\epsfxsize=0.66\textwidth
\begin{center}
\leavevmode
\epsffile{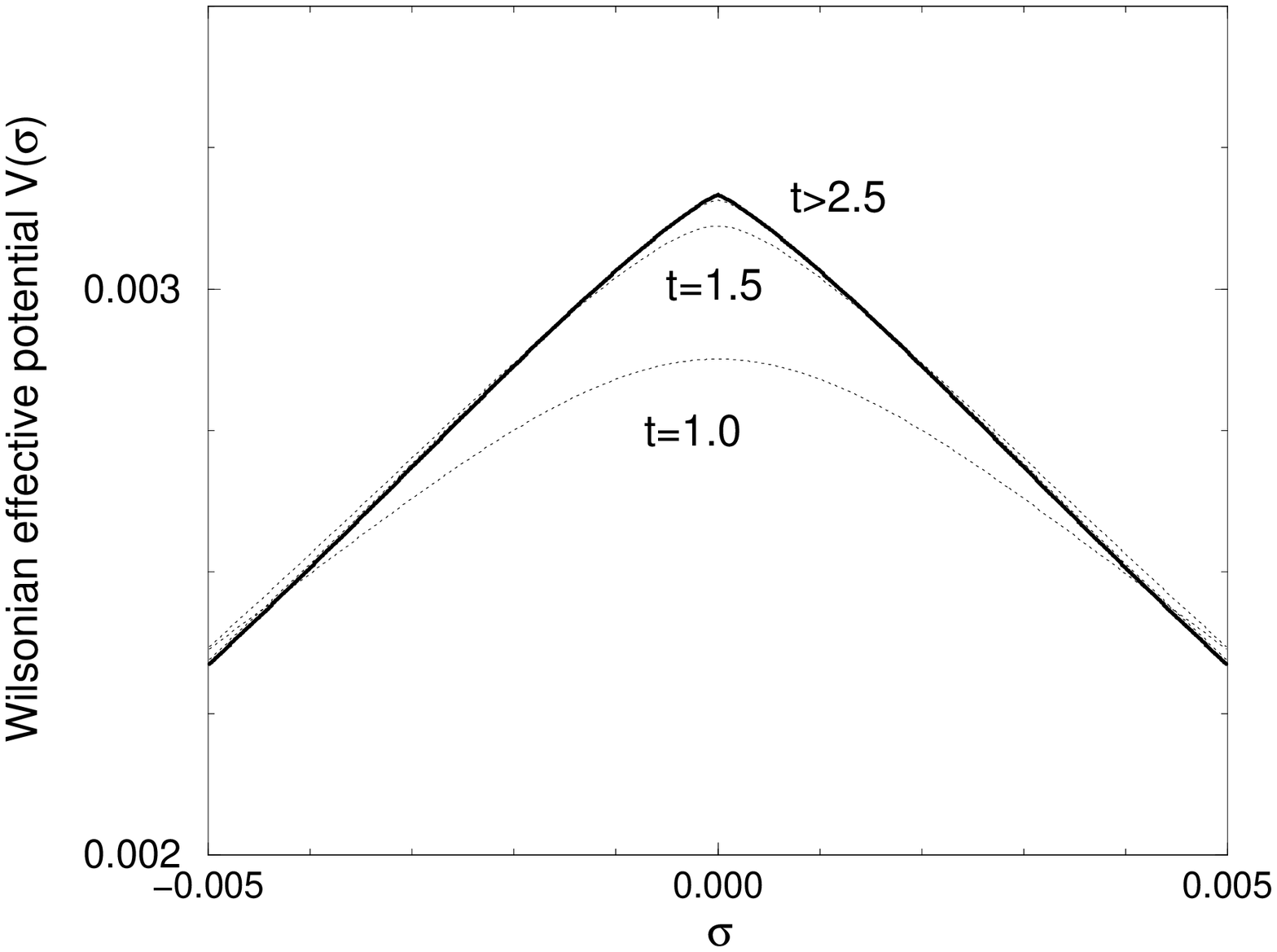}
\\
\parbox{130mm}{
Fig.2: The evolution of the Wilsonian effective potential
by the RGE (16) in the broken phase 
($G=0$, $\alpha=1.5~\alpha_{\mbox{cr}}$).
The effctive potential becomes non-analytic at the origin.
}
\end{center}
\end{figure}

Needless to say the approximation scheme for the non-perturbative 
RGE considered here is too clude to apply for the analysis of
QCD-like gauge theories. It is necessary to improve the approximation
so as to incorporate the running effect of the gauge coupling at
least. 
As the lowest order approximation to the RGE, 
the beta function for the gauge coupling 
may well be estimated by the perturbative one. 
Namely we solve the RGE given by eq.(16) coupled with the 
perturvative RGE for $\alpha$. 
This may be regarded somehow as a similar idea to the improved 
ladder approximation performed for the SDEs.
However it is a clear advantage for the Wilson RG approach to 
allow such improvement within the same framework of RG in contrast to 
the SD approach. 

Also we need to take the color group representation into consideration 
in the case of QCD-like gauge theories. 
Suppose the fermions form an $N$ dimensional representation.
Then the 1-loop correction given by (13) is multiplied by the factor
$N$. On the other hand we have to take care of the fierz transformation
for the color indices as well as the spinor indices to derive the
effective interaction as given in eq.(15). 
If we note the following identiy of the fierz transformation,
\be
\sum_{a}
\bar{\psi}_1\gamma_{\mu}T^a\psi_2 \bar{\psi}_3\gamma_{\nu}T^a\psi_4
=-\frac{C_2(R)}{4N}
\bar{\psi}_1\psi_4 \bar{\psi}_3\gamma_{\nu}\gamma_{\mu}\psi_2 
+\cdots,
\ee
then we may obtain the RGE similarly. 
The factor $N$ is found to be absorbed by the redefinition, 
$V \rightarrow NV$ and $\rho \rightarrow N^2 \rho$.
Consequently the RGE for the QCD-like gauge theories is given by
\be
\frac{\partial V}{\partial t}=
4V-6\rho V_{\rho} - \frac{1}{4\pi^2}\ln
\left[
1 + 2\rho(V_{\rho}-3\pi C_2(R)\alpha)^2
\right],
\ee
where the gauge coupling $\alpha(t)$ is now subject to the perturbative 
RGE. 
This RGE will be discussed in the comparison with the SDE in the
improved ladder approximation in the next section.

\vspace*{4mm}
\noindent
{\large \bf 4. The relation between the RGE and the SDE}
\vspace*{2mm}

In this section it is shown analytically that the RGE's, eq.(16)
and eq.(22) proposed in the previous section offer us the same 
effective masses as the SDE's in the (improved) ladder approximation 
do. 
Our strategy to this end is as follows.
First we give an analytical expression for the Wilsonian
effective potential, which is expected to be the solution of the RGE.
We define also the mass function giving the effective mass
introduced by eq.(20). 
The essential point in deriving these analytical expressions is to 
introduce the bilocal auxiliary field. 
Then it is shown that the mass function satisfies the (improved) ladder 
SDE as the stationary condition with respect to this auxiliary field.
On the other hand we also confirm that the effective potential given 
in this method indeed satisfies the RGE discussed before. 

The Wilsonian effective potential is given by integration of the
higher momentum modes. 
In the following discussions we use $\psi$, $\bar{\psi}$ and $A_{\mu}$ 
for the lower momentum modes, which are treated as zero modes in 
evaluation of the effective potential.
The higher momentum modes are expressed by $\chi_p$, $\bar{\chi}_p$ and
$a_{\mu p}$ respectively with index representing their momentum. 
For the time being also let us abbreviate the four fermi interaction 
$(\bar{\psi}i\gamma_5\psi)^2$ just for simplicity. 
In the approximation scheme considered in the previous section, we may 
give the following representation for the effective potential after 
integrating out $a_{\mu}$;
\be
e^{-\Omega V(\sigma; \Lambda)}
=e^{\Omega \frac{G}{2}\sigma^2}
\int {\cal D}\chi{\cal D}\bar{\chi}~e^{-S^{\chi}[\chi, \bar{\chi}; \sigma]}
\ee 
where $\sigma$ denotes the product of the lower (zero) momentum fermions
$\bar{\psi}\psi$ and $\Omega$ denotes the space-time volume.
$S^{\chi}$ is the action for the higher momentum fermions with
the effective 4-fermi interaction induced by the photon exchange,
which is given by
\be
S^{\chi}=
\int'_p \bar{\chi}_{-p}
\left[i\psla - \left(G+\frac{3g^2}{4p^2}\right)\sigma \right]\chi_p 
-\frac{1}{2}\int'_p\int'_{p'}\int'_k~
\left(G+\frac{3g^2}{4(k-p)^2}\right)
\bar{\chi}_{-p'}\chi_p \bar{\chi}_{-k}\chi_{k'},
\ee 
where $k'=k-p+p'$ by momentum conservation.
Here we introduced the simlpified notation for the momentum integration;
\be
\int'_p \equiv \int_{\Lambda<|p|<\Lambda_0}\frac{d^4 p}{(2\pi)^4}.
\ee

Now the effective 4-fermi interaction may be eliminated 
by adding the following term with a bilocal auxiliary field $S(p, p')$
\cite{bilocal};
\be
\delta S^{\chi, S}
=\frac{1}{2}\int'_p\int'_{p'}\int'_k~
\left(G+\frac{3g^2}{4(k-p)^2}\right)
\left(S(p,p')-\bar{\chi}_{-p'}\chi_p\right)
\left(S(k',k)-\bar{\chi}_{-k}\chi_{k'}\right).
\ee
Then we are able to integrate out the fermionic freedoms with higher
momentum, $\chi$ and $\bar{\chi}$.
After performing the integration we obtain the Wilsonian effective 
potential expressed by the path integration over the bilocal
auxiliary field $S$; 
\be
e^{-\Omega V(\sigma; \Lambda)}
=e^{\Omega \frac{G}{2}\sigma^2}
\int {\cal D}S~
e^{-S^{S}[S; \sigma]}.
\ee 
Here the effective action of the bilocal field is given by
\bea
S^S&=&
\frac{1}{2}\int'_p\int'_{p'}\int'_k~
\left(G+\frac{3g^2}{4(k-p)^2}\right)S(p,p')S(k',k) \nonumber \\
& &-\int'_p~\mbox{tr}\ln
\left[i\psla-\left(G+\frac{3g^2}{4p^2}\right)\sigma
+\int'_k~ \left(G+\frac{3g^2}{4(k-p)^2}\right)S(k,k)
\right].
\eea

The ladder approximation considered in the analyses of the SDE 
corresponds to the mean field approximation in the bilocal auxiliary
field method. Namely the effective potential in our approximation may 
be obtained by applying the saddle point method to the path integral 
given in eq.(27).
The stationary condition for the saddle point determines the expectation
value of the bilocal field, or the composite operator of the fermions
with higher momentum. Then the bilocal auxiliary field $S$ is now
reduced to the VEV depending on only the relative momentum,
\be
S(p,p') \rightarrow S(p)=\langle \bar{\chi}_{-p}\chi_p \rangle,
\ee
because of the translationary invariance.
Consequently we can obtain the following analytical expression of 
the Wilsonian effective potential, which is expected to satisfy the
RGE proposed in the preceeding section;
\be
V(\sigma;\Lambda)=
-\frac{G}{2}\sigma^2 
-\int'_p \mbox{tr}\ln\left[i\psla+\Sigma(p;\Lambda)\right]
+\frac{1}{2}\int'_p\int'_k~
\left(G+\frac{3g^2}{4(k-p)^2}\right)S(p)S(k),
\ee
where $\Sigma$ denotes the effective mass function for the higher
momentum modes, which is defined as
\be
\Sigma(p;\Lambda)=
-\left(G+\frac{3g^2}{4p^2}\right)\sigma
-\int'_k~
\left(G+\frac{3g^2}{4(k-p)^2}\right)S(k).
\ee
Note that this mass function depends on the lower momentum modes $\sigma$
as well as the IR cutoff $\Lambda$.
On the other hand the stationary condition with respect to $S(p)$ is 
found to be
\be
\Sigma(p;\Lambda)=
-\left(G+\frac{3g^2}{4p^2}\right)\sigma
-\int'_k~
\left(G+\frac{3g^2}{4(k-p)^2}\right)
\mbox{tr}\left(\frac{1}{i\ksla+\Sigma(k;\Lambda)}\right).
\ee
The mass function $\Sigma$ and also the auxiliary $S$ are  
formally given in terms of $\sigma$ by solving eq.(31) and eq.(32). 
Therefore the Wilsonian effective potential $V$ may be regarded as a
function of $\sigma$ and $\Lambda$.
This set of the eqautions will be found to interplay between
the ladder SDE (4) and the non-perturbative RGE in our approximation 
scheme (16).

First let us examine the effective mass defined by eq.(20).
By taking account of the stationary condition eq.(32), we may 
deduce from the eq.(30) the following relation, 
\bea
\frac{\partial V}{\partial \sigma}
&=&-G\sigma + \int'_k~\left(G+\frac{3g^2}{4k^2}\right)
\mbox{tr}\left(\frac{1}{i\ksla+\Sigma(k;\Lambda)}\right)\nn \\
&=&-G\sigma + \frac{1}{2\pi^2}
\int_{\Lambda}^{\Lambda_0}dk~k^3~
\left(G+\frac{3g^2}{4k^2}\right)
\frac{\Sigma(k; \Lambda)}{k^2+\Sigma(k;\Lambda)^2}.
\eea
Therefore the effective mass is found to be given in terms of 
$\Sigma$ defined above by
\be
\meff \equiv \lim_{\Lambda \rightarrow 0}
\left.
\frac{\partial V}{\partial \sigma}(\sigma;\Lambda)
\right|_{\sigma=0}
=\lim_{p \rightarrow 0} \lim_{\Lambda \rightarrow 0}
\left. \Sigma(p; \Lambda)\right|_{\sigma=0}.
\ee
On the other hand it is readily seen by setting $\sigma=0$
that the mass function $\Sigma(p; \Lambda)$ indeed satisfies the 
ladder SDE given by eq.(4) in the IR limit of $\Lambda=0$. 
Thus it is verified
that the effective mass obtained by solving the ladder SDE is also
derived from the Wilosnian effective potential given by (30).

Next we shall confirm that the effective potential also satisfies
the RGE given by eq.(16). Again by using the stationary condition (32),
we may obtain
\be
\Lambda\frac{\partial V}{\partial \Lambda}=
\Lambda \int \frac{d^4p}{(2\pi)^4}~\delta(|p|-\Lambda)
\mbox{tr}\ln[i\psla + \Sigma(p; \Lambda)].
\ee
Here if we set $|p|=\Lambda$ in eq.(32),
then it is written down after performing the angle integration
of the momentum as
\be
\Sigma(|p|=\Lambda; \Lambda)=
-\left(G+\frac{3g^2}{4\Lambda^2}\right)\sigma
-\frac{1}{2\pi^2}\int_{\Lambda}^{\Lambda_0}dk~k^3
\left(G+\frac{3g^2}{4k^2}\right)
\frac{\Sigma(k; \Lambda)}{k^2+\Sigma(k;\Lambda)^2}.
\ee
Therefore the variation of the effective 
potential $V(\sigma; \Lambda)$ given by eq.(35) may be represented 
in terms of the potential itself by using eq.(33) and is found to be
\be
\Lambda\frac{\partial V}{\partial \Lambda}=
\frac{1}{2\pi^2}\Lambda^4
\ln\left[
\Lambda^2 + \left(V_{\sigma}-\frac{3g^2}{4\Lambda^2}\sigma\right)^2
\right].
\ee
Thus it is seen that the effective potential derived through the above 
discussion indeed satisfies the RGE given in (16) after considering the
canonical scaling transfaormation.

The SDE in the improved ladder approximation also can be related to the
RGE by the similar arguments given so far. 
In this case it is enough to replace the gauge coupling appearing 
in the analytical expression of the Wilsonian effective potential to 
the running one; 
$g^2 \rightarrow C_2(R)\bar{g}^2(\mbox{max}(p^2, k^2))$.
Then it is now obvious that the effective mass function $\Sigma$
at $\sigma=0$ satisfies the improved ladder SDE (7) in the IR
limit. 
On the other hand the variation of the effective potential
is found to be given by
\be
\Lambda\frac{\partial V}{\partial \Lambda}=
\frac{1}{2\pi^2}\Lambda^4
\ln\left[
\Lambda^2 + \left(V_{\sigma}-
\frac{3C_2(R)\bar{g}^2(\Lambda^2)}{4\Lambda^2}\sigma\right)^2
\right].
\ee
Here it should be noted that this variation may be given in terms of the
potential owing to the special choice of the momentum dependence of the 
running gauge coupling in (6). Since the argument of the running gauge 
coupling appearing in eq.(38) is simply given by the cutoff $\Lambda$, 
we may realize that the effective potential is the solution of the coupled 
equations of the non-perturbative RGE given by eq.(22) and the 
RGE for the gauge coupling. 
Thus it has been shown that the improved ladder approximation, 
which looks rather artificial in the SD approach, turns out
to be the naive improvement by evaluating the gauge beta function 
perturbatively in the non-perturbative renormalization group 
framework.

In this section we discussed the relation between the non-perturbative RGE
and the (improved) ladder SDE via the Wilsonian effective potential given by
using the bilocal auxiliary field. 
However we should note that the above argument is rather formal. 
As a matter of fact the ladder SDE (4) is known to
have many solutions corresponding to the unstable states. 
Therefore the solution of the RGE for the effective potential might not be
uniquely determined at the singular point in the broken phase.
We have not cleared this point yet. 
It has been also known that the unstable solutions of the SDE are related
to the appearance of the tachyon pole for the bilocal auxiliary field 
\cite{bilocal}. 
More careful analyses with taking the bound states into
consideration would be necessary to understand the unstable solutions
in the SDE's in the RG framework.

\vspace*{4mm}
\noindent
{\large \bf 5. The RG equations with the collective field}
\vspace*{2mm}

In section 3 we discussed the dynamical mass of the fermion appears
through non-analytic behavior of the effective potential in infrared.
Therefore in order to see the spontaneous chiral symmetry breaking,
it seems to be necessary to deal with infinetely many operators, 
or to solve the partial differential equation in practice, 
since otherwise such non-analyticity does not come out. 
However it turns out to be rather difficult to calculate the effective 
mass in a good accuracy with this manner.  
Hence we might suppose that the Wilson RG approach is 
unsuitable for evaluation of the order parameters, while it is indeed
quite efficient to analyze the critical dynamics. 

On the other hand the non-analytic behavior of the effective potential
was found to be related with the infrared singularity caused by the 
massless fermions.
However the fermions must acquire thier mass dynamically in the broken 
phase and, therefore, the system should be free from the infrared 
singularity. 
Then how can we treat the mass to be generated in the RG framework
with keeping it's chiral symmetry?
Actually it will be seen that this problem is resolved by introducing proper 
new operators into the theory space. 

Our method is to extend the theory space by introducing the 
collective coordinate corresponding the order parameter 
$\langle \bar{\psi}\psi\rangle$.
Concretely we simply introduce the collective field, which 
denotes $\phi$, as an auxiliary field in the original path
integration.
In this section we discuss how the order parameters may be
evaluated from the non-perturbative RGE defined in the extended 
space in the relation with the SDE's.
It will be shown that the order parameters are 
calculated quite precisely by dealing with only several operators,
that is, by truncating the power series of the effective potential
by a few terms. 
Thus this method will be seen to be much more effective than the 
naive Wilson RG in the present case.

In the followings we shall consider the QED with the 4-fermi 
interaction by abbreviating $(\bar{\psi}i\gamma_5\psi)^2$ part
for simplicity. 
Then we may rewrite the partition function as
\be
Z=\int {\cal D}\psi{\cal D}\bar{\psi}{\cal D}A_{\mu}{\cal D}\phi
e^{-S_0[\psi, \bar{\psi}, A_{\mu}, \phi]},
\ee
where the bare action $S_0$ is given by
\be
S_0=\int d^4x~
\bar{\psi}(\dsla + g\Asla)\psi 
-\frac{G}{2}(\bar{\psi}\psi)^2+\frac{1}{2}(\phi-y\bar{\psi}\psi)^2 
+\frac{1}{4}F_{\mu\nu}^2+\frac{1}{2\alpha}(\partial_{\mu}A_{\mu})^2,
\ee
which is invariant under the discrete chiral transformation;
\be
\psi \rightarrow \gamma_5\psi, \hspace{7mm}
\bar{\psi} \rightarrow -\bar{\psi}\gamma_5, \hspace{7mm}
\phi \rightarrow -\phi.
\ee
Here if we choose the Yukawa coupling $y$ so as to elimimate 
the four-fermi interaction, this action may be regarded as the 
gauged Yukawa system imposed the so-called compositeness 
condition, which was first examined by using the one loop RGE 
by Bardeen, Hill and Lindner \cite{BHL}. 
However it should be stressed that the collective coordinates 
are introduced not for the purpose of making the
perturbative treatment possible by eliminating the 4-fermi
interactions, but to avoid the infrared singurality by reflecting
the dynamically generated fermion mass faithfully to the RGE.
\footnote{In this sense our formulation is a sort of the 
``environmentally friendly renormalization group'', which has been
proposed by O'Connor and Stephens \cite{OS}.
}
Note that we may treat the system equally even in the case that 
the four-fermi interaction is absent in the bare action.
It is also noted that the resultant order parmeters should not 
depend on the choice of the Yukawa coupling introduced in the bare 
action. 

First we shall derive the RGE for this system in the same approximation 
scheme considered in section 3. Then the Wilsonian effective
action at the general scale of $\Lambda$ will be given by
\be
\Seff=\int d^4x~
\bar{\psi}(\dsla + g\Asla)\psi 
+ U(\phi, \sigma; \Lambda)
+ \frac{1}{4}F_{\mu\nu}^2 +\frac{1}{2\alpha}(\partial_{\mu}A_{\mu})^2,
\ee
where the effective potential $U$ is kept invariant under the chiral
transformation (41). It is noted that the collective field $\phi$ does 
not appear as a propagating mode through the evolution
due to the local potential approximation. 
Actually it is found that the contributions of these field to the 
radiative corrections are totally ignored in the approximation 
applied here. 
\footnote{
For the NJL model this scheme is nothing but the large N 
leading approximation.
}
Therefore the RGE is derived with just same argument done in section 
3 and is found to be
\be
\frac{\partial U}{\partial t}=
4U-3\sigma U_{\sigma}
-\phi U_{\phi}
-\frac{1}{4\pi^2}\ln
\left[
1+\left(U_{\sigma}-3\pi\alpha\sigma \right)^2
\right].
\ee 
Thus the structure of the RGE incorporating the collective field
has been reduced to be of the same form as the RGE considered so far
due to this simple approximation. 
Nevertheless this formulation will be found to have a great 
advantage compared to the former one. 

First let us explain the way to evaluate the order parameters
with this RGE and the numerical results.
If we expand the effective potential $U(\phi, \sigma; t)$ into
a polinomial with respect to $\sigma$ as
\be
U(\phi, \sigma; t)=
U^{(0)}(\phi; t)+U^{(1)}(\phi; t)\sigma
+\frac{1}{2}U^{(2)}(\phi; t)\sigma^2 + \cdots,
\ee
then eq.(43) is reduced to the RGE's for the functions 
$U^{(i)}(\phi; t)$. In the chiral symmetry broke phase 
the potential $U^{(0)}(\phi)$ will be found to show it's 
non-trivial minimum at a certain scale under evolution,
\footnote{
The Wilsonian effective potential is shown to coincide
in the infrared limit 
with the conventional effective potential defined by the 
Legendre transformation, which must be convex \cite{convexity}.
However the RG given eq.(43) leads us to evolution into the concave
one due to the naive approximation adopted here.
This behavior is related with the divergence of the RG flows 
at a finite scale disscussed in section 3.
}
and the collective field acquires the vacuum expectation value 
$\langle \phi \rangle$.
Shortly it will be shown that $\langle \phi \rangle$ is
related with the chiral condensate by 
\be
\langle \bar{\psi}\psi \rangle =
\frac{1}{y} \langle \phi \rangle,
\ee
which is independent of the Yukawa coupling $y$. 
The point of this method is to solve the RGE at the absolute minimum
of this potential $U^{(0)}(\phi)$. 
\footnote{
There found two absolute minima correspondingly to the discrete 
chiral symmetry. If the small bare mass is introduced, the absolute
minima is uniquely determined.
}
There the fermion propagates with the effective mass appearing
through the linear term with respect to $\sigma$ as
\be
\meff(t)=U^{(1)}(\langle \phi \rangle; t).
\ee
Thus we may avoid the problem of the infrared singularity.
The RG flows are subject to the canonical scaling at the scale 
lower than the effective mass.
Indeed the effective mass $\meff=\meff(t\rightarrow \infty)$ 
obtained in this method will be shown to coincide with the results 
in the ladder SD approach. 

In the practical analysis we may well truncate the series given
by eq.(44) at a certain order. The results are found to converge 
quite rapidly with increasing the number of the couplings.
In Fig.3 the contour of the effective fermion mass obtained by this
method is presented in the space of the bare couplings 
$(G, \alpha)$ compared with the corresponding results
by numerical analysis of the ladder SDE. It is
clearly seen that they coincide with each other perfectly as
is expected from the analytical study done in the previous section.
The proof of this coincidence based on the RGE with the collective
field (43) will be also given shortly.
We show also the effective mass generated in the pure gauge theories
compared with the SD results in Fig.4. Notice that the effective
mass of the theory in the very vicinity to the criticality is
evaluated enough precisely. These observations may convince us that the
RG method is quite efficient in analyses of the composite order 
parameters as well as the critical behavior in the 
dynamical chiral symmetry breaking. 

\begin{figure}[hbt]
\epsfxsize=0.5\textwidth
\leavevmode
\epsffile{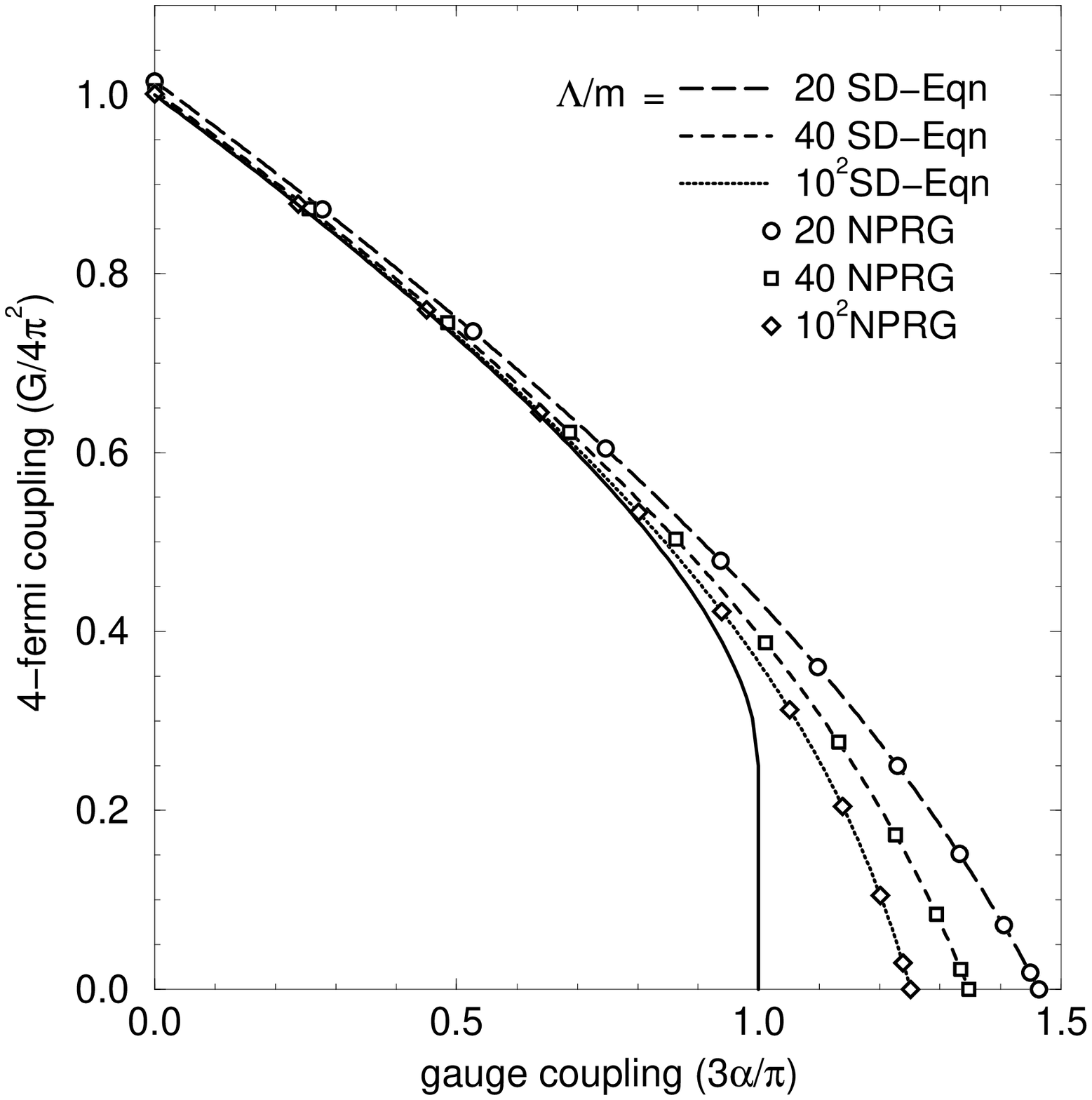}
\epsfxsize=0.5\textwidth
\leavevmode
\epsffile{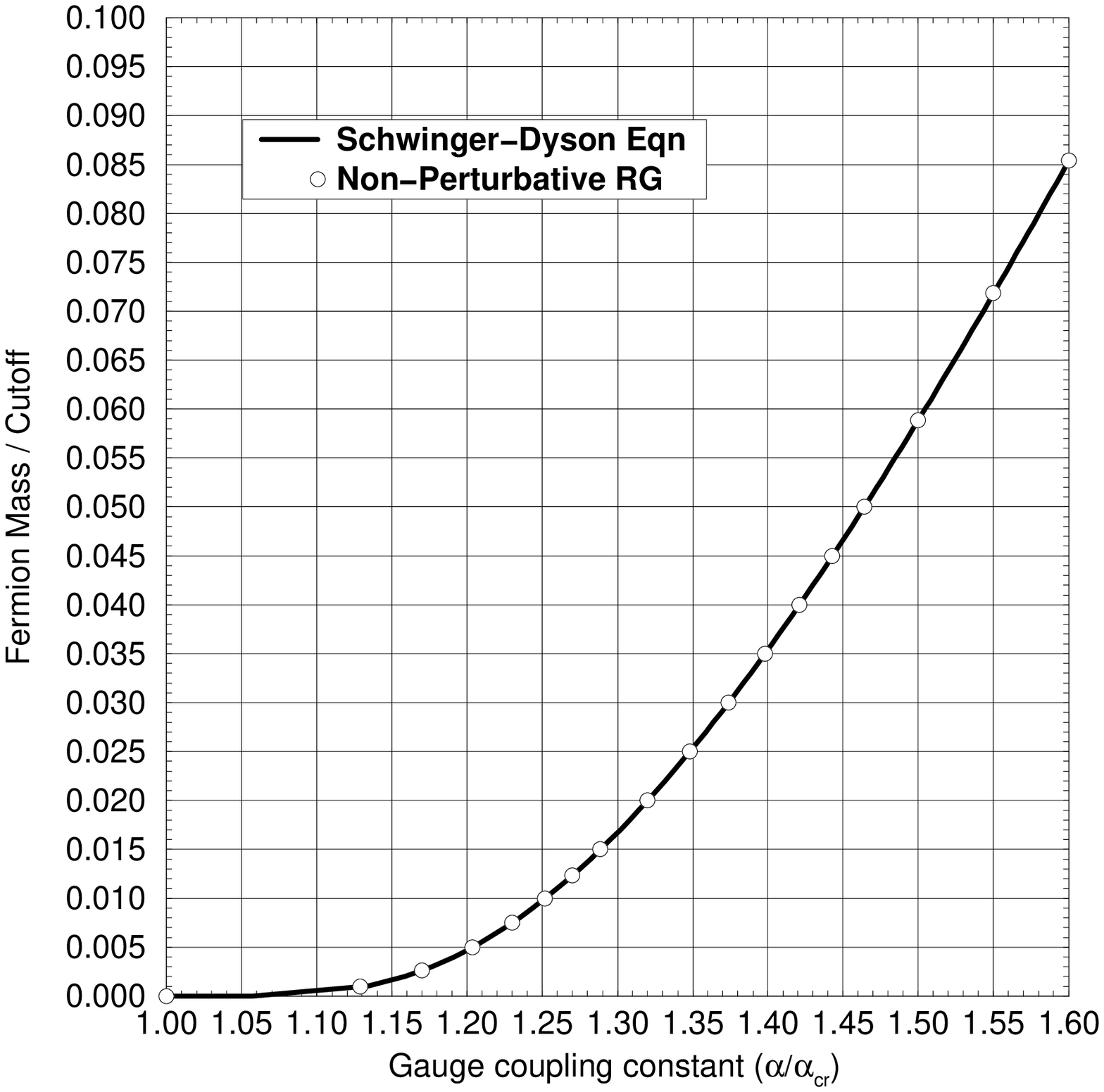}
\begin{center}
\parbox{75mm}{
Fig.3: Contour of the effective mass in the $(G, \alpha)$ plane.
The results obtained by solving the non-perturbative RGE coincide
with the ladder SD results.
}
\hspace*{5mm}
\parbox{75mm}{
Fig.4: The effective mass in strongly coupled gauge theories; $(G=0)$.
}
\end{center}
\end{figure}

In the primitive approximation scheme applied here, which reproduces
the results obatined by solving the (improved) ladder SDE, 
any radiative corrections by the collective field has been ignored.
However the collective field introduced just as an 
auxiliary in the bare action becomes the propagating degree 
of freedom as an bound state with evolution.
Therefore it will be significant to improve the approximation
so that the colective field is incoporated as the dynamical 
variable.
\footnote{The non-perturbative RGE's for the Higgs-Yukawa theories
have been examined for study of the triviality mass
bound in the standard model \cite{yukawaRG}. 
}
For this purpose the analysis in the scheme beyond the local 
potential approximation will be necessary.

Now we shall show along the argument done in the previous
section that the order parameters, i.e. the effective mass and the chiral
condensate, are really independent of the Yukawa coupling and also 
accord with the SD results. 
For the time being $\phi$ and $\sigma$ denote the
lower momentum modes, which are treated as zero modes.
Since the collective field does not contribute to
the radiative correction at all,
we may immediately written down the Wilsonian effective potential
$U(\phi, \sigma; \Lambda)$ by introducing the bilocal auxiliary
field;
\bea
U(\phi, \sigma; \Lambda)
&=&-\frac{G-y^2}{2}\sigma^2 -y \phi\sigma +\frac{1}{2}\phi^2
-\int'_p \mbox{tr}\ln\left[i\psla+\Sigma(p;\Lambda)\right] \nn\\
& &+\frac{1}{2}\int'_p\int'_k~
\left(G-y^2+\frac{3g^2}{4(k-p)^2}\right)S(p)S(k),
\eea
where the effective mass function $\Sigma$ denotes
\be
\Sigma(p;\Lambda)=
-\left(G-y^2+\frac{3g^2}{4p^2}\right)\sigma-y\phi
-\int'_k~
\left(G-y^2+\frac{3g^2}{4(k-p)^2}\right)S(k).
\ee
Also the stationary condition is found to be given by
\be
\Sigma(p;\Lambda)=
-\left(G-y^2+\frac{3g^2}{4p^2}\right)\sigma -y\phi
-\int'_k~
\left(G-y^2+\frac{3g^2}{4(k-p)^2}\right)
\mbox{tr}\left(\frac{1}{i\ksla+\Sigma(k;\Lambda)}\right).
\ee
We note that the mass function depends on not only $\sigma$ but 
also $\phi$ in turn.
It is staightforward to see that the effective potential 
$U(\phi, \sigma; \Lambda)$ defined by these equations 
satisfies the non-perturbative RGE (43) indeed. 
However the relation to the ladder SDE is not obvious due to the
presence of the collective field $\phi$.

By solving the RGE (43) we may obtain the effective potential 
$U(\phi,\sigma)$ as the infared limit. 
However the effective mass is defined through the effective 
potential in terms of the fermion, 
$V(\sigma; \Lambda)$ as is given by eq.(20). 
In order to derive this effective potential from $U(\phi, \sigma)$
we have to integrate out the collective zero mode $\phi$;
\be
e^{-\Omega V(\sigma)}=\int d\phi e^{-\Omega U(\phi,\sigma)}.
\ee  
However we may well evaluate this integral by the saddle
point method at the absolute minimum of the potential $U$
in the infrared (infinite volume) limit. 
Then $\phi$ is given by a function in terms of $\sigma$ 
after solving the stationary condition of $U$;
\be
\frac{\partial U}{\partial \phi}=
-y\sigma +\phi +y \int\frac{d^4p}{(2\pi)^4}
\left(\mbox{tr}\frac{1}{i\psla + \Sigma(p)}\right)=0.
\ee
Let us represent the formal solution for this equation by 
$\phi^*(\sigma)$. By using this the effective potential in terms 
of the fermions is given by $V(\sigma)=U(\phi^*(\sigma),\sigma)$. 
On the other hand the minimum of the potential $U^{(0)}$, which
was denoted by $\langle \phi \rangle$ above is nothing but 
\be
\langle \phi \rangle = \phi^*(\sigma=0) 
=-y\int\frac{d^4p}{(2\pi)^4}
\left. \left(\mbox{tr}\frac{1}{i\psla + \Sigma(p)}\right)
\right|_{\sigma=0}.
\ee

First we shall consider the effective mass function $\Sigma(p)$ 
which should be subject to the ladder SDE in the infrared limit. 
Note that the collective coordinate $\phi$ in the stationary 
condition (49) is now given by $\phi^*(\sigma)$. Therefore by
setting $\sigma=0$, we may obtain the effective mass function as
\be
\Sigma(p)=-y\langle \phi \rangle 
+\int\frac{d^4k}{(2\pi)^4}
\left(G-y^2+\frac{3g^2}{4(k-p)^2}\right)
\left. \left(\mbox{tr}\frac{1}{i\ksla + \Sigma(k)}\right)
\right|_{\sigma=0}.
\ee
After substituting the expressoin for $\langle \phi \rangle$
given by eq.(52) into this equation, it is readily seen that the 
effective mass indeed satisfies the ladder SDE and, therefore, is
independent of the free parameter $y$. 

Now it is seen that the order parameters defined by
eq.(45) and eq.(46) in the context of the RG
are just the same quantities given by the ladder SDE,
and, therefore, are independent of $y$.
Coincidence of the effective mass is shown by the relation
\be
\left. \frac{dV(\sigma)}{d\sigma}\right|_{\sigma=0}
=\lim_{t \rightarrow \infty}
\left. \frac{\partial U(\phi, \sigma; t)}{\partial \sigma}
\right|_{\phi=\langle \phi \rangle, \sigma=0}
=\lim_{t \rightarrow \infty}U^{(1)}(\langle \phi \rangle; t)
=\meff.
\ee
We note also that this effective mass is the momentum independent
part of the the mass function given by eq.(53); $\meff=\Sigma(0)$.
From eq.(52) we see that the chiral condensate is given by 
$\langle \bar{\psi}\psi \rangle = -\langle \phi \rangle /y $.
Since the mass function has been shown to be independent of $y$, 
the chiral condensate is found to be so. 
Thus the non-perturbative RGE incorporating the collective field (43)
also has been shown to offer the order parameters 
in the ladder approximation.
 
This method is found to be efficient equally in the analysis of
the dynamical chiral symmetry breaking in QCD-like gauge theories.
The order parameters may be evaluated by solving the RGE given by
eq.(43) with the running gauge coupling imposed a proper IR
cutoff. 
The bare cuttoff scale $\Lambda_0$ should be large enough
compared to the dynamical scale of the theory.
Now it would be expected that this method using the 
non-perturbative RGE should reproduce the identical 
results with those obtained by using the improved ladder SDE, 
as long as we atopt the same IR cutoff to the running gauge coupling. 
In Fig.5 the results of the dynamical mass of quarks 
$\meff=\Sigma(0)$ as well as the chiral condensate 
$\langle \bar{\psi}\psi \rangle$ obtained by the both method are shown
in the case of SU(3) QCD with three triplet massless quarks.
From this figure we may realize the efficiency of this RG method,
with the collective filed, since the results converge quite
rapidly against truncation of the potential given by (44). 
In the practical calculation, the scale of $\Lambda_{\mbox{QCD}}$ is
set to 490MeV, which has been known to offer the $\pi$ decay constant of
the experimental value $f_{\pi}=$94MeV \cite{SDQCD}.  
The chiral condensates are renormalized at 1GeV.

\begin{figure}[hbt]
\epsfxsize=0.66\textwidth
\begin{center}
\leavevmode
\epsffile{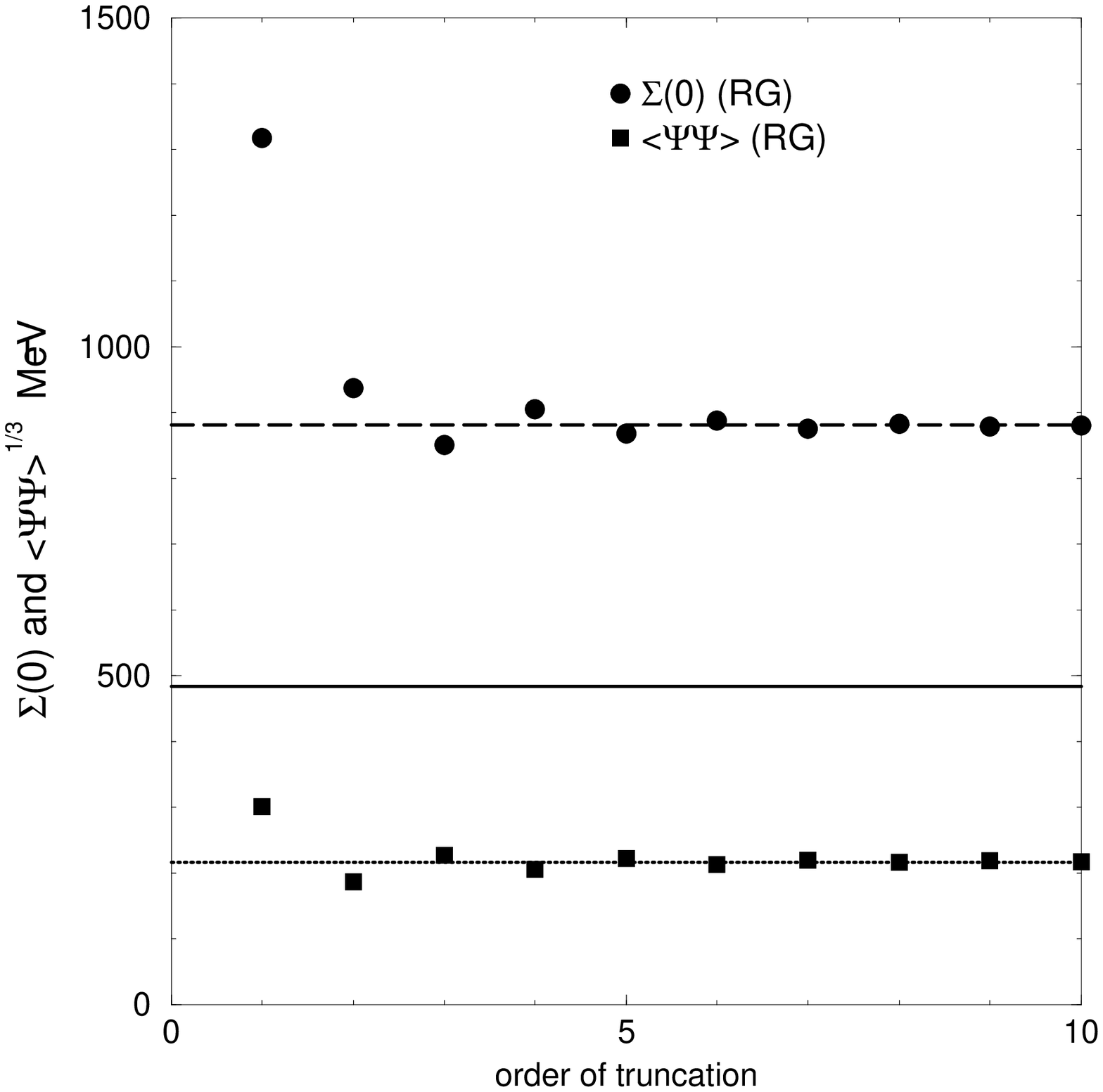}
\\
\parbox{130mm}{
Fig.5: The effective mass and the chiral condensate obtained by the RG
method and their truncation dependence is shown. The values for these
order parameters obtained by the ladder SD analyses are given by the 
dashed line and the dotted line respectively.
}
\end{center}
\end{figure}

\vspace*{4mm}
\noindent
{\large \bf 6. Discussions}
\vspace*{2mm}

In this paper we have proposed non-perturbative RGE's by which
we can analyze the order parameters for the chiral symmetry 
breaking dynamically induced by strong gauge interactions. 
The RGE was derived in the simple approximation scheme, which 
exactly accords with the ladder approximation mostly 
used in the SD approach. 
Moreover the so-called improved ladder approximation performed
in the SD framework has been understood as the very natural and 
simple extension of the RGE.
The spontaneous symmetry breaking is found to appear through 
the non-analyticity of the infrared effective potential in the 
naive Wilson RG picture. We have proposed also the scheme of the 
non-perturbative RG extended so as to incorporate the collective 
coodinate corresponding to the composite order parameter in order 
to avoid the infrared singularity. 
This extended RGE is quite efficient in the practical evaluation of
the order parameters, which was demonstrated by the calculation
in the case of the massless QCD. 

Throughout of this paper we have been discussing in a quite simple
approximation scheme in order to make clear relation with the SD
approach. Now it has been found that any improvement of this
approximation enable us to go beyond the analyses by the (improved)
ladder approximation. As far as the critical dynamics is concerned,
it has been achieved so as to incorporate the non-ladder diagrams in the 
previous works \cite{ccbqed,RGfourfermi}. 
We should stress that the gauge dependence, which has been a serious 
problem in the SD approach \cite{SDdifficulty}, 
remarkably disappears by taking account of the non-ladder diagrams
\cite{ccbqed}.  
Therefore it seems to be important to extend the analyses for the 
order parameters presented in this paper to the better approximation 
scheme including the non-ladder corrections. 

Another interesting problem in the non-perturbative RG study 
would be the bound states and the composite particles.
In this paper we have introduced the collective field
$\phi$ representing the fermion composite operator.
It is naturally expected that this field is closely related 
to the composite particle. 
Recently the appearance of the composite particle has
been vigorously studied in the non-perturbative 
RG framework \cite{quarkmeson}. 
It would be extremely fascinating problem to describe the 
transition of the effective degree of freedom in QCD, quarks
to hadrons, in the Wilson RG picture.

\end{document}